\documentclass[%
 reprint,
nofootinbib,
floatfix,
superscriptaddress]{revtex4-1}
\usepackage[english]{babel}
\usepackage{siunitx}
\usepackage{dcolumn}
\usepackage{BOONDOX-calo}
\usepackage[caption=false]{subfig}
\usepackage{floatrow}
\usepackage{amsfonts}
\usepackage{microtype}
\usepackage{amsmath}
\usepackage{mathtools}
\usepackage{tabularx}
\usepackage{appendix}
\usepackage{chngcntr}
\usepackage{graphicx}% Include figure files
\usepackage{dcolumn}% Align table columns on decimal point
\usepackage{bm}% bold math
\usepackage[colorlinks=true, allcolors=blue, pdfencoding=auto]{hyperref}
\usepackage[multiple]{footmisc}
\usepackage{fnpct}

\usepackage[Symbolsmallscale]{upgreek}
\DeclareMathAlphabet\mathbfcal{OMS}{cmsy}{b}{n}

\addto\extrasenglish{%
}

\addto\extrasenglish{%
}

\addto\extrasenglish{%
}

\addto\extrasenglish{%
}

\addto\extrasenglish{%
}
\addto\extrasenglish{%
}

\usepackage[capitalize,nameinlink]{cleveref}
\usepackage[integrals]{wasysym}
\creflabelformat{equation}{#2#1#3}
\crefname{section}{Sec.}{Secs.}
\DeclareMathAlphabet{\mathpzc}{OT1}{pzc}{m}{it}

\begin{document}

%\preprint{APS/123-QED}

\title{Deep Inelastic Scattering Cross Section Uncertainties in Tau Neutrino Appearance Measurements}

\author{Tetiana Kozynets}%
 \email{tetiana.kozynets@nbi.ku.dk}
\affiliation{Niels Bohr Institute, University of Copenhagen, Jagtvej 155A, 2200 Copenhagen, Denmark}
\author{Thomas Stuttard}
\affiliation{Niels Bohr Institute, University of Copenhagen, Jagtvej 155A, 2200 Copenhagen, Denmark}
\author{D. Jason Koskinen}
\affiliation{Niels Bohr Institute, University of Copenhagen, Jagtvej 155A, 2200 Copenhagen, Denmark}

\date{\today}

\begin{abstract}
In neutrino experiments sensitive to multiple flavors, the analyzers may be presented with a choice of treating the uncertainties on the respective cross sections in a correlated or an uncorrelated manner. This study focuses on the charged current deep inelastic scattering (CC DIS) channel in experiments sensitive to both muon and tau neutrinos. We evaluate the ratio of the leading-order $\nu_{\tau}$ and $\nu_{\mu}$ cross sections and derive its uncertainty from the underlying parton distribution functions (PDFs).  We find that, for neutrino energies above 5 GeV, the PDF-driven uncertainty on the cross section ratio is less than 3\%, with a larger (\SIrange[range-phrase=--,range-units=single]{2}{30}{}\%) variation seen in antineutrinos at energies below \SI{10}{GeV}. These results suggest that for atmospheric tau neutrino appearance analyses, the uncertainties in $\nu_{\tau}$ and $\nu_{\mu}$ DIS cross sections should be coupled, while separate treatment for the two flavors may be warranted in long-baseline experiments with an antineutrino beam. We further explore the role of the invariant hadronic mass threshold defining the onset of the DIS regime. We argue that its impact may be incorporated only if it is applied to both DIS and resonance cross sections, and if the correlations with other DIS and resonance cross section parameters are taken into account.

\end{abstract}

\maketitle

\section{Introduction}\label{sec:intro}

The uncertainty on the charged current deep inelastic scattering (CC DIS) neutrino cross section is an important systematic parameter in experimental analyses involving neutrino detection at energies of a few GeV and above. The scenario when a detector is sensitive to all three neutrino flavors, which applies to e.g. atmospheric neutrino oscillation experiments, is of particular interest, as it raises the question of correlations between the uncertainties on the cross sections of the different flavors. The flavor dependence of the DIS cross section appears in terms proportional to $m_{\ell}^2 / E_{\nu}^2$ or $m_{\ell}^2 / (E_{\nu} M_N)$, where $E_{\nu}$ is the incident neutrino energy, $M_N$ is the mass of the target nucleon, and $m_{\ell}$ is the mass of the produced charged lepton \cite{Kretzer:2002fr,Albright:1974ts,Jeong:2010nt,Ansari:2020xne}. These terms are typically neglected for $\nu_{e}$ and $\nu_{\mu}$ cross sections due to the smallness of the electron and the muon masses relative to the neutrino energies in the DIS regime. However, the mass of the tau lepton ($m_{\tau} \simeq\,$\SI{1.7}{GeV}) is comparable to the $\mathcal{O}$(\SIrange[range-phrase=--,range-units=single]{1}{10}{GeV}) energies relevant for atmospheric and accelerator neutrino oscillation measurements. Naively, this could result in the conclusion that the $\nu_{\tau}$ CC DIS cross section uncertainty should be decoupled from the $\nu_{\mu}$ and $\nu_e$ uncertainties in an experimental analysis, i.e., by letting the energy-dependent $\nu_{\tau}$ uncertainty vary independently from those of other flavors. The present study challenges this hypothesis by computing the $\nu_{\tau} / \nu_{\mu}$ CC DIS cross section ratio at the leading order in perturbation theory and estimating its uncertainty. We achieve the latter by propagating the uncertainties on the parton distribution functions (PDFs) into the structure functions entering the DIS cross section, which we evaluate numerically. We additionally study the dependence of the cross section ratio on the threshold mass of the hadronic final state defining the transition between the resonant (RES) and the deep inelastic scattering. The outcomes of these tests are meant to guide neutrino experiments in defining the scope of the cross section systematic uncertainties necessary for neutrino oscillation analyses in the DIS regime. This concerns, in particular, the experiments such as IceCube-DeepCore \cite{IceCube:2017lak,IceCube:2019dqi,IceCubeCollaboration:2023wtb,IceCube:2024xjj} and KM3NeT \cite{KM3NeT:2021ozk}, whose $\nu_{\mu}$ and $\nu_{\tau}$ event selections are dominated by DIS interactions, as well as  the upcoming IceCube-Upgrade \cite{Ishihara:2019aao,IceCube:2023ins} and DUNE \cite{DUNE:2020lwj,DeGouvea:2019kea}, which cover the transition region between RES and DIS regimes (also referred to as ``shallow inelastic scattering'').

\section{The Charged Current DIS Cross Section}\label{sec:dis_formalism}

Neutrino-nucleon ($\nu N$) charged current deep inelastic scattering can be characterized through the kinematic variables $x$ and $y$, representing the fraction of the nucleon momentum carried away by the struck quark (also called the Bjorken scaling variable) and the inelasticity of the interaction, respectively. For a neutrino $\nu_{\ell}$ (antineutrino $\bar{\nu}_{\ell}$), the double-differential cross section with respect to these two variables reads \cite{Kretzer:2002fr,Albright:1974ts,Jeong:2010nt,Ansari:2020xne}:

\begin{align}
\frac{\mathrm{d}^2 \sigma^{\nu_{\ell} N (\bar{\nu}_{\ell} N)}}{\mathrm{d}x \mathrm{d}y} = & \frac{G_F^2 M_N E_{\nu}}{\pi (1 + Q^2 / M_{W}^2)^2} \Bigg\{  \left(y^2 x + \frac{m_{\ell}^2 y}{2 E_{\nu} M_N} \right) F_1 \nonumber \\& + \left[ \left( 1 - \frac{m_{\ell}^2}{4 E_{\nu}^2} \right) - \left( 1 + \frac{M_N x}{2 E_{\nu}}  \right) y \right] F_2 \nonumber 
\\&\pm \left[xy \left( 1 - \frac{y}{2} \right)  - \frac{m_{\ell}^2 y}{4E_{\nu}M_N}\right]F_3 \label{eq:dsigma_dx_dy}\\ &+ \frac{m_{\ell}^2 (m_{\ell}^2 + Q^2)}{4E_{\nu}^2 M_N^2 x}F_4 - \frac{m_{\ell}^2}{E_{\nu}M_N} F_5 \Bigg\}\nonumber.
\end{align}
Here, $Q^2 \equiv 2 M_N E_{\nu} x y$ is the momentum transfer squared, $G_F$ is the Fermi constant, and $F_i \equiv F_i (x, Q^2)$, $i=1..5$, are the structure functions of the nucleon. The ``$+$'' sign in the coefficient multiplying $F_3$ corresponds to neutrinos, and the ``$-$'' sign -- to antineutrinos. While the terms proportional to either $m_{\ell}^2 / E_{\nu}^2$ or $m_{\ell}^2 / (E_{\nu} M_N)$ enter the coefficients multiplying all of the 5 structure functions, the contributions due to $F_4$ and $F_5$ are effectively negligible when $m_{\ell}^2 \ll (E_{\nu} M_N)$. For this reason, they are typically not included in the calculation of the $\nu_{e}$ or $\nu_{\mu}$ DIS cross sections (see e.g. \cite{Candido:2023utz}) but become relevant in the case of $\nu_{\tau}$ \cite{Kretzer:2002fr,Jeong:2010nt,Ansari:2020xne}. We include these structure functions explicitly when computing both $\nu_{\mu}$ and $\nu_{\tau}$ cross sections, as described in \cref{sec:PDF_to_structure_functions}. 

All of the structure functions $F_i$ have an underlying dependence on the experimentally determined PDFs, whose uncertainties ultimately propagate to the DIS cross section in \cref{eq:dsigma_dx_dy}. At large $Q^2$ ($Q^2 \gtrsim 1\,\mathrm{GeV}^2$, where the strong coupling constant $\alpha_\mathrm{s}$ is smaller than unity \cite{Workman:2022ynf,Deur:2016tte}), the path from the PDFs to the structure functions lies though perturbation theory. In this context, the leading-order (LO), next-to-leading order (NLO), and next-to-next-to-leading order (NNLO) perturbative QCD calculations of the absolute $\nu_{\tau}$ cross sections have been performed in literature \cite{Kretzer:2002fr,Jeong:2010nt,Ansari:2020xne,Ansari:2021cao,Jeong:2023hwe}. The contribution of each scattering diagram of order $n$ is weighted by $\alpha_{\mathrm{s}}^n(Q^2)$, which is monotonically decreasing with growing $Q^2$. This implies that the difference between the results of the leading-order and the higher-order calculations is the largest at the lowest $Q^2$ and, consequently, $E_{\nu}$\footnote{At $E_{\nu} =\,$\SI{10}{GeV}, the difference between the $\nu_{\tau}$ CC DIS cross sections evaluated at LO vs NLO is $\sim$10\% \cite{Kretzer:2002fr}.}. Furthermore, the low-$Q^2$ region is susceptible to non-perturbative effects such as the target mass corrections (concerning the masses of the target nucleon and the struck quark) and the dynamical higher twist effects (concerning the interactions the of struck quark with the surrounding quarks)   \cite{Ansari:2020xne,Ansari:2021cao}. In this study, we evaluate the structure functions and the inclusive DIS cross sections at the leading order to capture the dominant impact of the PDF uncertainty propagation. Additionally, we take into account the corrections for the nucleon mass $(M_N \approx 0.938\,\mathrm{GeV})$ and the mass of the final-state charm quark ($m_c \approx\,$\SI{1.27}{GeV}), as these effects introduce a small deviation of the $F_4$ structure function from 0 at leading order \cite{Kretzer:2002fr}. Finally, we perform all of our calculations for an isoscalar target, assuming that the nuclear medium effects apply equally to $\nu_{\mu}$ and $\nu_{\tau}$ scattering and do not influence the cross section ratio\footnote{The effects such the $W$-boson interactions with the mesonic cloud and the nuclear shadowing were shown to introduce $\mathcal{O}$(tens of \%) shifts to the $^{56}$Fe structure functions in regions of intermediate $x$ (\SIrange[range-phrase=--,range-units=single]{0.1}{0.5}{}) and $Q^2$ (\SIrange[range-phrase=--,range-units=single]{2}{20}{}\,$\mathrm{GeV}^2$) \cite{Ansari:2021cao}. This is comparable to the difference in the structure functions due to the calculation order, e.g. LO vs. NLO (NNLO) \cite{Kretzer:2002fr,Candido:2023utz}. The impact of nuclear effects increases with the mass number and is therefore expected to be smaller for nuclei such as $^{16}$O (part of the IceCube and KM3NeT water target) or $^{40}$Ar (DUNE target).}.

\section{Structure functions at leading order}\label{sec:PDF_to_structure_functions}

Throughout this study, we work with the conventional mass basis for quarks, such that the $u$, $c$, $t$ mass eigenstates coincide with the weak interaction eigenstates, while the $d$, $s$, $b$ mass eigenstates are related to the weak eigenstates $d'$, $s'$, and $b'$ through the CKM matrix \cite{Workman:2022ynf}: 
\begin{equation}
    \begin{pmatrix}
    d' \\  s' \\  b' \end{pmatrix} = \begin{pmatrix}
        V_{ud} & V_{us} & V_{ub} \\
        V_{cd} & V_{cs} & V_{cb} \\
        V_{td} & V_{ts} & V_{tb}
    \end{pmatrix}\begin{pmatrix}
    d \\  s \\  b \end{pmatrix}.
\end{equation}
Additionally, we operate in the fixed-flavor number scheme with $n_f = 4$ quark flavors, i.e., consider only the first two generations \footnote{The density of the sea charm quark relative to that of the sea strange quark is given in \cref{fig:sea_charm_density}. We find that the ultimate impact of including the sea charm into the $\nu_{\tau} / \nu_{\mu}$ cross section ratio calculations is near-negligible (see \cref{fig:charm_tmc_impact_on_xsec_ratio}).}. From the charge conservation constraints, a neutrino $\nu_l$ corresponding to the lepton flavor $l$ can engage in CC interactions with the $d'$, $s'$, $\bar{u}$, or $\bar{c}$ quarks inside a nucleon. Such interactions are mediated by the $W^{+}$ boson and result in the production of the $u$, $c$, $\bar{d}'$, and $\bar{s}'$ final-state quarks, accordingly. Analogously, a CC scattering of an antineutrino $\bar{\nu}_l$ can result in the $\bar{d}' \to \bar{u}$, $\bar{s}' \to \bar{c}$, $u \to d$, and $c \to s'$ conversions at the $W^{-}$ vertex.  

As the PDFs measured by the experimental collaborations probe the proton structure, we will use $\mathbcal{q}(x, Q^2)$ to denote the density of the quark $q$ inside a proton. These PDFs can be used to compute the structure functions $F_i\,(i = 1..5)$ for both neutrino-proton and neutrino-neutron scattering by applying the isospin symmetry argument. At leading order and ignoring the target nucleon and the charm mass corrections (i.e., assuming $M_N = 0$, $m_c = 0$), the $F_2$ and $F_3$ structure functions for the $\nu p$ scattering are \cite{Candido:2023utz,Ansari:2020xne}
\begin{subequations}
\begin{align}
    & F_2^{\nu p}(x,Q^2)\Bigg|_{\substack{M_N = 0 \\ m_c = 0}} = 2x[\mathbcal{d}' + \mathbcal{s}' + \bar{\mathbcal{u}} + \bar{\mathbcal{c}}](x, Q^2);\label{eq:f2_no_tmc}\\
    &F_3^{\nu p}(x,Q^2)\Bigg|_{\substack{M_N = 0 \\ m_c = 0}} = 2[\mathbcal{d}' + \mathbcal{s}' - \bar{\mathbcal{u}} - \bar{\mathbcal{c}}](x, Q^2)\label{eq:f3_no_tmc},
\end{align}
\label{eq:f2_f3_no_tmc}
\end{subequations}
where $\mathbcal{d}' = |V_{ud}|^2 \mathbcal{d} + |V_{us}|^2 \mathbcal{s}$ and $\mathbcal{s}' = |V_{cd}|^2 \mathbcal{d} + |V_{cs}|^2 \mathbcal{s}$. For the $\bar{\nu} p$ scattering, the PDFs for all quarks in \cref{eq:f2_f3_no_tmc} are replaced by those of their weak doublet partners, e.g., $\mathbcal{d}' \to \mathbcal{u}$, while for the $\nu n$ scattering, the replacements are made only for the $u$ and $d$ quarks (see \cite{Ansari:2020xne} for complete expressions).
Furthermore, in the considered massless approximation at LO, $F_4(x, Q^2) = 0$, while $F_1$ and $F_5$ follow from the Callan-Gross \cite{Callan:1969uq} and the Albright-Jarlskog \cite{Albright:1974ts} relations:
\begin{subequations}
    \begin{align}
        & \Big[F_1(x, Q^2) - \frac{F_2(x, Q^2)}{2x}\Big]\Bigg|_{\substack{M_N = 0 \\ m_c = 0}} = 0; \\
        & \Big[F_5(x, Q^2) - \frac{F_2(x, Q^2)}{2x}\Big]\Bigg|_{\substack{M_N = 0 \\ m_c = 0}} = 0.
    \end{align}
\label{eq:cg_aj_relations_f1_f5_at_lo}
\end{subequations}
To account for the non-zero nucleon mass $M_N$, the Bjorken $x$ variable in the quark PDFs has to be replaced by the Nachtmann variable $\eta$, defined as \cite{Kretzer:2002fr,Nachtmann:1973mr}
\begin{equation}
    \frac{1}{\eta} = \frac{1}{2x} + \sqrt{\frac{1}{4x^2} + \frac{M_N^2}{Q^2}}.
\label{eq:nachtmann_eta_correction_quark_mass}
\end{equation}
The Nachtmann variable is further corrected by the final-state quark mass $m_q$:
\begin{equation}
    \eta \to \bar{\eta}_q \equiv \frac{\eta}{\lambda_q} = {\eta} \left(\frac{Q^2}{Q^2 + m_q}\right)^{-1},
\end{equation}
where we assume only the charm quark to be massive ($m_{q, q \neq c} = 0$). Further, we define 
\begin{equation}
    \rho^2 = 1 + \left(\frac{2M_N x}{Q}\right)^2, 
\label{eq:rho_definition}
\end{equation}
following \cite{Kretzer:2002fr, Ansari:2020xne}. Then, the final leading-order expressions for the structure functions corrected by the masses of the nucleon and the final-state charm quark become\footnote{The expressions in \cref{eq:proton_Fi_at_LO_corrected_by_MN_mc} are given in a collinear approximation \cite{Aivazis:1993kh}, i.e., where the momenta of the struck parton and the parent nucleon are aligned in the infinite momentum frame. This approximation can be relaxed to include the parton transverse momentum \cite{Ellis:1982cd}, which leads to the full target mass corrections as in \cite{Kretzer:2003iu,Schienbein:2007gr}. While these expressions would lead to more accurate absolute cross sections, our goal is to estimate the magnitude of the PDF-driven uncertainty on the $\nu_{\tau}$/$\nu_{\mu}$ cross section ratio, and we deem the collinear approximation plausible for this purpose.}\footnote{The ultimate impact of the $M_N$ and $m_c$ corrections on the $\nu_{\tau} / \nu_{\mu}$ cross section ratio is shown in \cref{fig:charm_tmc_impact_on_xsec_ratio}.} \cite{Kretzer:2002fr}:

\begin{widetext} 
\begin{subequations}
\begin{align}
    & F_1^{\nu p}(x, Q^2) =  \mathbcal{d}'(\eta, Q^2) + \mathbcal{s}'(\bar{\eta}_c, Q^2) + \bar{\mathbcal{u}}(\eta, Q^2) + \bar{\mathbcal{c}}(\eta, Q^2);\\
        & F_2^{\nu p}(x, Q^2) = \frac{2x}{\rho^2} \Big[\mathbcal{d}'(\eta, Q^2) + \frac{1}{\lambda_c}\mathbcal{s}'(\bar{\eta}_c, Q^2) + \bar{\mathbcal{u}}(\eta, Q^2) + \bar{\mathbcal{c}}(\eta, Q^2)\Big];\\
        & F_3^{\nu p}(x, Q^2) = \frac{2}{\rho} \Big[\mathbcal{d}'(\eta, Q^2) + \mathbcal{s}'(\bar{\eta}_c, Q^2) - \bar{\mathbcal{u}}(\eta, Q^2) - \bar{\mathbcal{c}}(\eta, Q^2)\Big]; \\
        & F_4^{\nu p}(x, Q^2) =  \Big[\frac{1 - \rho^2}{2\rho^2}\Big ] \cdot \Big[\mathbcal{d}'(\eta, Q^2) + \bar{\mathbcal{u}}(\eta, Q^2) + \bar{\mathbcal{c}}(\eta, Q^2)\Big] + \Big[ \frac{(1 - \rho)^2}{2\lambda_c\rho^2} + \frac{1 - \rho}{\rho}\Big]\mathbcal{s}'(\bar{\eta}_c, Q^2);\\
        & F_5^{\nu p}(x, Q^2) =  \frac{1}{\rho^2}\Big[\mathbcal{d}'(\eta, Q^2) + \bar{\mathbcal{u}}(\eta, Q^2) + \bar{\mathbcal{c}}(\eta, Q^2)\Big] + \Big[ \frac{1}{\rho} - \frac{\rho - 1}{\lambda_c \rho^2}\Big]\mathbcal{s}'(\bar{\eta}_c, Q^2).
\end{align}
\label{eq:proton_Fi_at_LO_corrected_by_MN_mc}
\end{subequations}
\end{widetext}
We note that with $M_N \to 0$, \cref{eq:rho_definition} gives $\rho = 1$, which recovers $F_4 = 0$ and the relations from \cref{eq:cg_aj_relations_f1_f5_at_lo}. The isospin symmetry arguments can again be applied to  \cref{eq:proton_Fi_at_LO_corrected_by_MN_mc} to derive the neutron structure functions $F_i^{\nu n}$. From here, the structure functions for an isoscalar target $A$ are obtained as
 \begin{equation}
     F_i^{\nu A} = \frac{F_i^{\nu p} + F_i^{\nu n}}{2}.
 \end{equation}
 
\begin{figure*}[htb!]
\includegraphics[width=\textwidth]{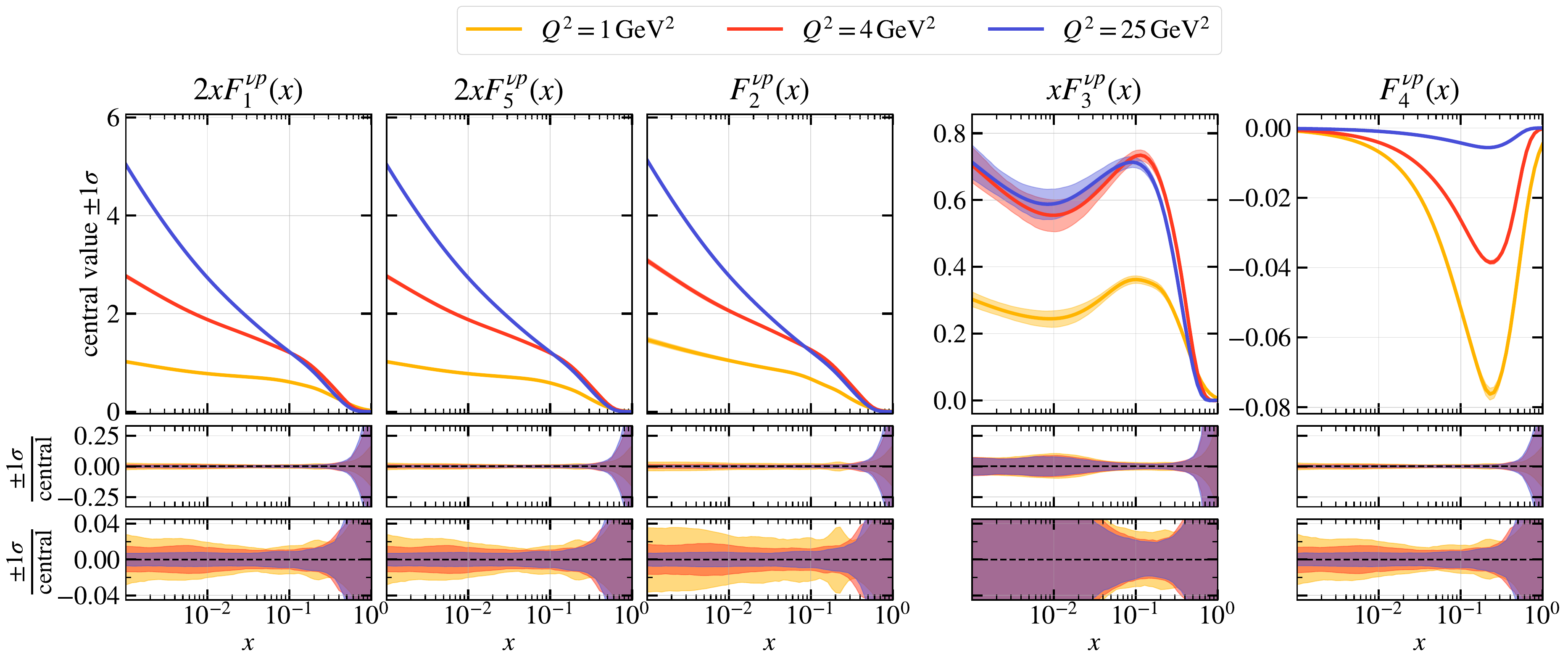}
\caption{Structure functions $F_i(x)$ evaluated at fixed $Q^2$ at LO and using the NNPDF4.0 PDF grids \cite{NNPDF:2021njg}. In the upper panels, the shaded bands represent the 68\% ($1\sigma$) confidence level, while the middle and the bottom panels show the ratio of the $1\sigma$ band widths to the central value. At $Q^2 < Q_{\mathrm{min}}^2 = (1.65\,\mathrm{GeV})^2$, the MSTW extrapolation scheme \cite{Martin:2009iq, Buckley:2014ana} is used.}
\label{fig:struc_funcs_fixed_Q2}
\end{figure*}

\begin{figure*}[htb!]
\includegraphics[width=\textwidth]{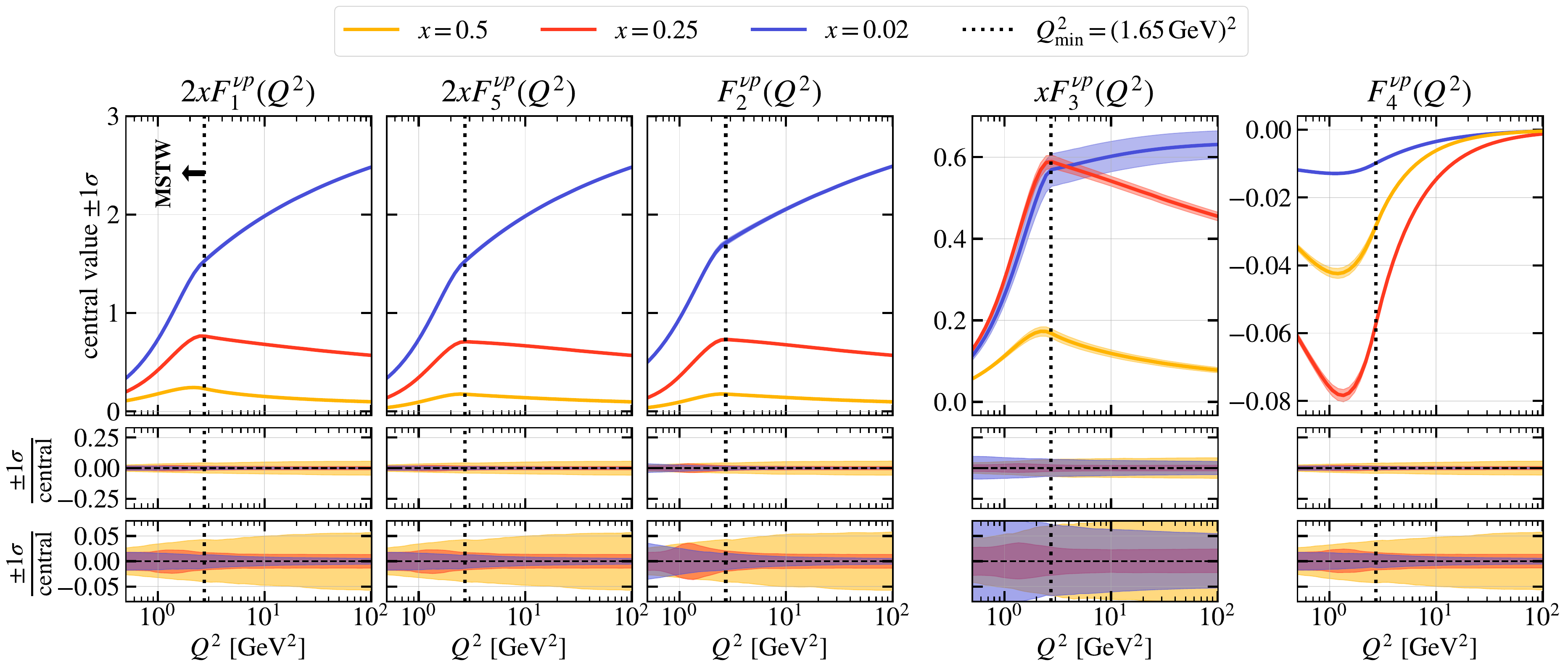}
\caption{Structure functions $F_i(Q^2)$ evaluated at fixed $x$ at LO (for details, see \cref{fig:struc_funcs_fixed_Q2}). The vertical dotted black line denotes the minimum $Q^2$ value of the NNPDF4.0 PDF grids, below which the MSTW extrapolation scheme is used \cite{Martin:2009iq, Buckley:2014ana}.}
\label{fig:struc_funcs_fixed_x}
\end{figure*}

 In \cref{fig:struc_funcs_fixed_Q2}, we present the $F_i^{\nu p}(x)$ structure functions evaluated at fixed $Q^2$, while \cref{fig:struc_funcs_fixed_x} shows $F_i^{\nu p}(Q^2)$ at fixed $x$. We are using the NNPDF4.0 PDF set \cite{NNPDF:2021njg} fitted at NNLO and applied at LO according to \cref{eq:proton_Fi_at_LO_corrected_by_MN_mc}, as done in the benchmark tests of \cite{Candido:2023utz}. The PDFs are accessed through the LHAPDF library \cite{Buckley:2014ana} interfaced with the \textsc{PDFFlow} Python module \cite{Carrazza:2020qwu}. As the NNPDF4.0 grids are provided and valid down to the minimum momentum transfer value of $Q_{\mathrm{min}} =\,$\SI{1.65}{GeV}, we use the Martin-Stirling-Thorne-Watt (MSTW) scheme\footnote{This is the default extrapolation scheme for all PDF grids in the recent versions of the LHAPDF library (see Sec. 3.4.2 in \cite{Buckley:2014ana}).} built into \textsc{PDFFlow} for extrapolation to the low-$Q^2$ region \cite{Martin:2009iq}. In addition to the central values of the PDFs, the NNPDF set includes 1000 PDF ``replicas''  obtained by fitting 1000 individual neural networks to the Monte Carlo replicas of the original data \cite{Forte:2002fg,NNPDF:2017mvq}. We use these replicas to construct the 68\% ($1\sigma$) uncertainty contours for the structure functions, which are shown in \cref{fig:struc_funcs_fixed_Q2,fig:struc_funcs_fixed_x} alongside their central values. The distribution of the structure function replicas ultimately yields the corresponding distribution of the total DIS cross sections for $\nu_{\mu}$ and $\nu_{\tau}$ flavors and allows us to derive the PDF-driven uncertainty on their ratio, as described in \cref{sec:inclusive_xsecs}.

\section{Inclusive DIS cross sections and $\nu_{\tau} / \nu_{\mu}$ cross section ratio at LO}\label{sec:inclusive_xsecs}

\begin{figure}[htb!]
\includegraphics[width=0.8\textwidth]{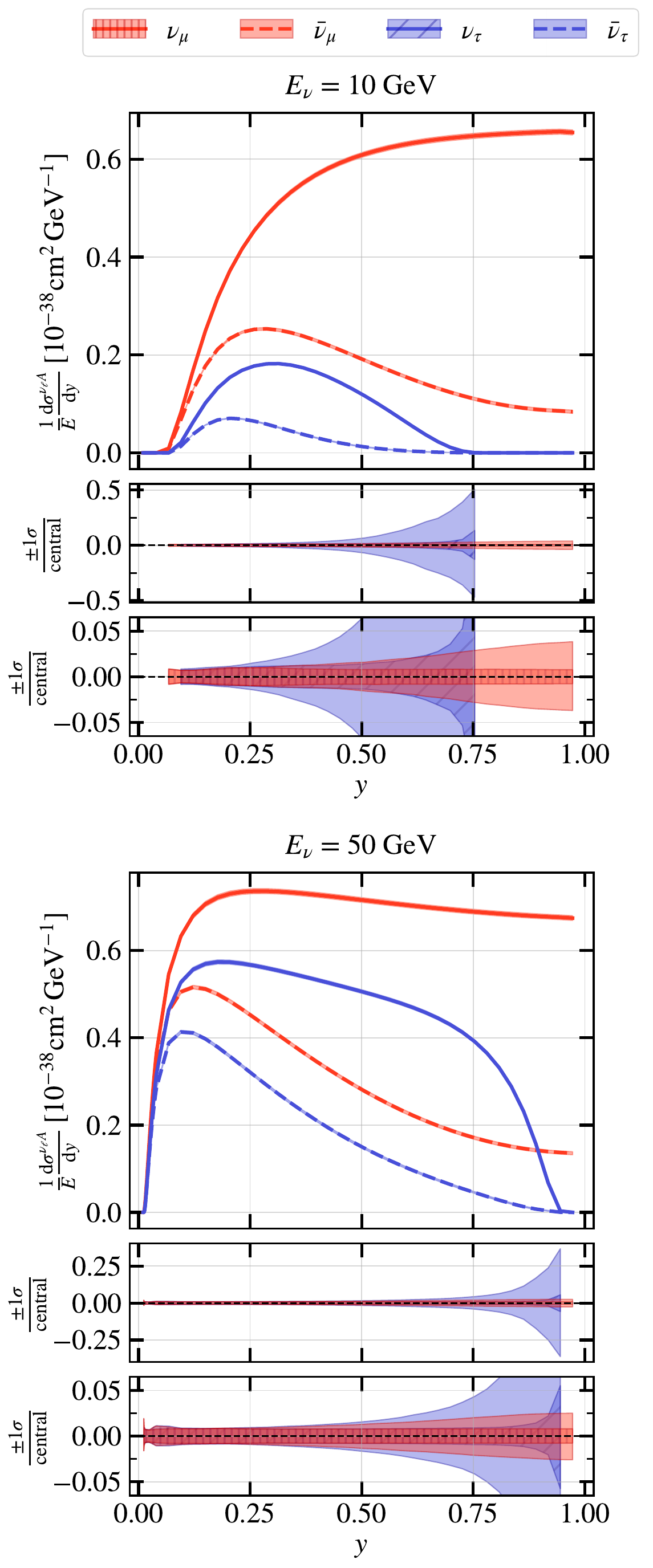}
\caption{Differential CC DIS cross sections of $\nu_{\mu}, \bar{\nu}_{\mu}$ (red curves) and $\nu_{\tau}, \bar{\nu}_{\tau}$ (blue curves) with an isoscalar target $A$, obtained at leading order using the structure functions from \cref{eq:proton_Fi_at_LO_corrected_by_MN_mc}. The details on the PDF set and the $1\sigma$ uncertainty range (shaded bands) derivation are given in \cref{sec:PDF_to_structure_functions}.}
\label{fig:diff_xsec}
\end{figure}

\begin{figure}[htb!]
\includegraphics[width=0.8\textwidth]{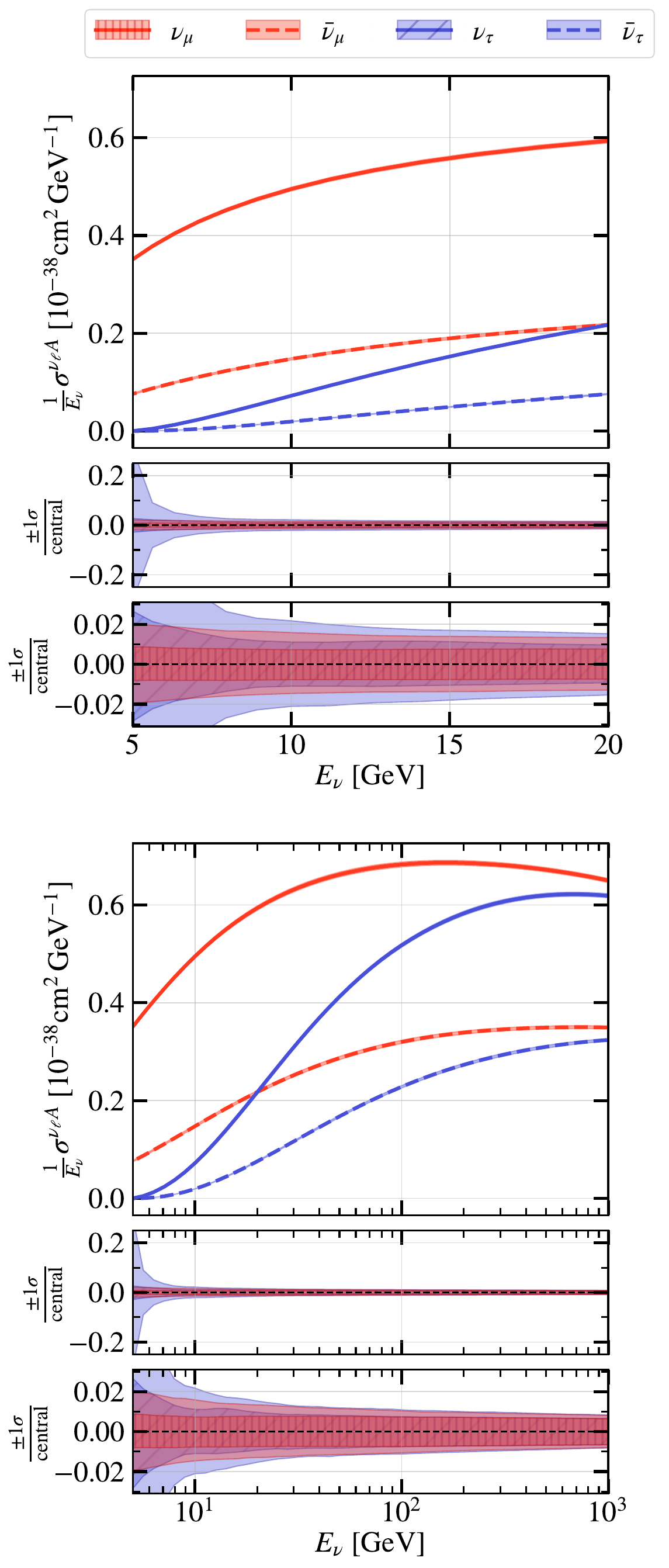}
\caption{Total CC DIS cross sections of $\nu_{\mu}, \bar{\nu}_{\mu}$ (red curves) and $\nu_{\tau}, \bar{\nu}_{\tau}$ (blue curves) with an isoscalar target $A$, obtained at leading order by integrating \cref{eq:dsigma_dy} over the inelasticity range and using the structure functions from \cref{eq:proton_Fi_at_LO_corrected_by_MN_mc}. The details on the PDF set and the $1\sigma$ uncertainty range (shaded bands) derivation are given in \cref{sec:PDF_to_structure_functions}. The top and the bottom panel respectively show the low energy range (\SIrange[range-phrase=--,range-units=single]{5}{20}{GeV}) and the full energy range (\SIrange[range-phrase=--,range-units=single]{5}{1000}{GeV}).}
\label{fig:total_xsec}
\end{figure}

The double-differential cross section $\frac{\mathrm{d}^2 \sigma}{\mathrm{d}x \mathrm{d}y}$ from \cref{eq:dsigma_dx_dy} can alternatively be expressed as a function of $y$ and $Q^2$ through the change of variables:
\begin{equation}
    \frac{\mathrm{d}^2 \sigma}{\mathrm{d}y \mathrm{d}Q^2} = \frac{\mathrm{d}^2 \sigma}{\mathrm{d}x \mathrm{d}y} \frac{\partial x}{\partial Q^2} = \frac{1}{2M_N E_{\nu} y} \, \frac{\mathrm{d}^2 \sigma}{\mathrm{d}x \mathrm{d}y}.  
\end{equation}
Then, the single-differential cross section with respect to inelasticity can be found by integrating over $Q^2$:
\begin{equation}
    \frac{\mathrm{d}\sigma}{\mathrm{d}y} = \int_{Q^2_{\mathrm{min}}}^{Q^2_{\mathrm{max}}}\frac{\mathrm{d}^2 \sigma}{\mathrm{d}y \mathrm{d}Q^2} \mathrm{d}Q^2.
\label{eq:dsigma_dy}
\end{equation}
We reproduce the $Q^2$ integration limits from \cite{Jeong:2010nt} for completeness:
\begin{subequations}
\begin{align}
    & Q^2_{\mathrm{min}} = 2E_{\nu}^2(1 - \epsilon)(1 - y) - m_{\ell}^2;\\
    &Q^2_{\mathrm{max}} = 2M_N E y + M_N^2 - W_{\mathrm{min}}^2,
\end{align}
\label{eq:q2_integration_limits}
\end{subequations}

\noindent where $\epsilon = \sqrt{1 - \frac{m_{\ell}^2}{((1 - y) E_{\nu})^2}}$ and $W_{\mathrm{min}}$ is the minimum invariant mass of the hadronic final state required to classify an interaction as deep inelastic as opposed to resonant scattering. We use $W_{\mathrm{min}} =\,$\SI{1.4}{GeV} throughout this study and further test the impact of the $W_{\mathrm{min}}$ choice in \cref{sec:pdf_wmin_impact}.
In \cref{fig:diff_xsec}, we show the result of the $\frac{\mathrm{d}\sigma}{\mathrm{d}y}$ calculation through \cref{eq:dsigma_dy}, including the 68\% confidence limit ($1\sigma$) uncertainties obtained from the 1000 NNPDF replicas. \Cref{fig:total_xsec} further gives the total cross sections obtained by integrating $\frac{\mathrm{d}\sigma}{\mathrm{d}y}$ over the allowed inelasticity range\footnote{These results can be compared to those from \cite{Kretzer:2002fr,Jeong:2010nt} obtained at NLO.}. 
 
We observe that the relative cross section uncertainty is larger at higher inelasticities and lower energies, with both of these effects being more prominent for $\nu_{\tau}$ than $\nu_{\mu}$. This could be explained by the larger relative contributions of the $F_4$ and $F_5$ structure functions to the DIS cross section in case of the $\nu_{\tau}$ scattering, with their coefficients being proportional to $y$ and inversely proportional to $E_{\nu}$. We also find that the uncertainty on both single-differential and total cross sections is larger for antineutrinos than for neutrinos, which can be attributed to the PDF uncertainty cancellation being dependent on the sign with which $F_3$ structure function enters \cref{eq:dsigma_dx_dy}.

\begin{figure*}[htb!]
\includegraphics[width=0.8\textwidth]{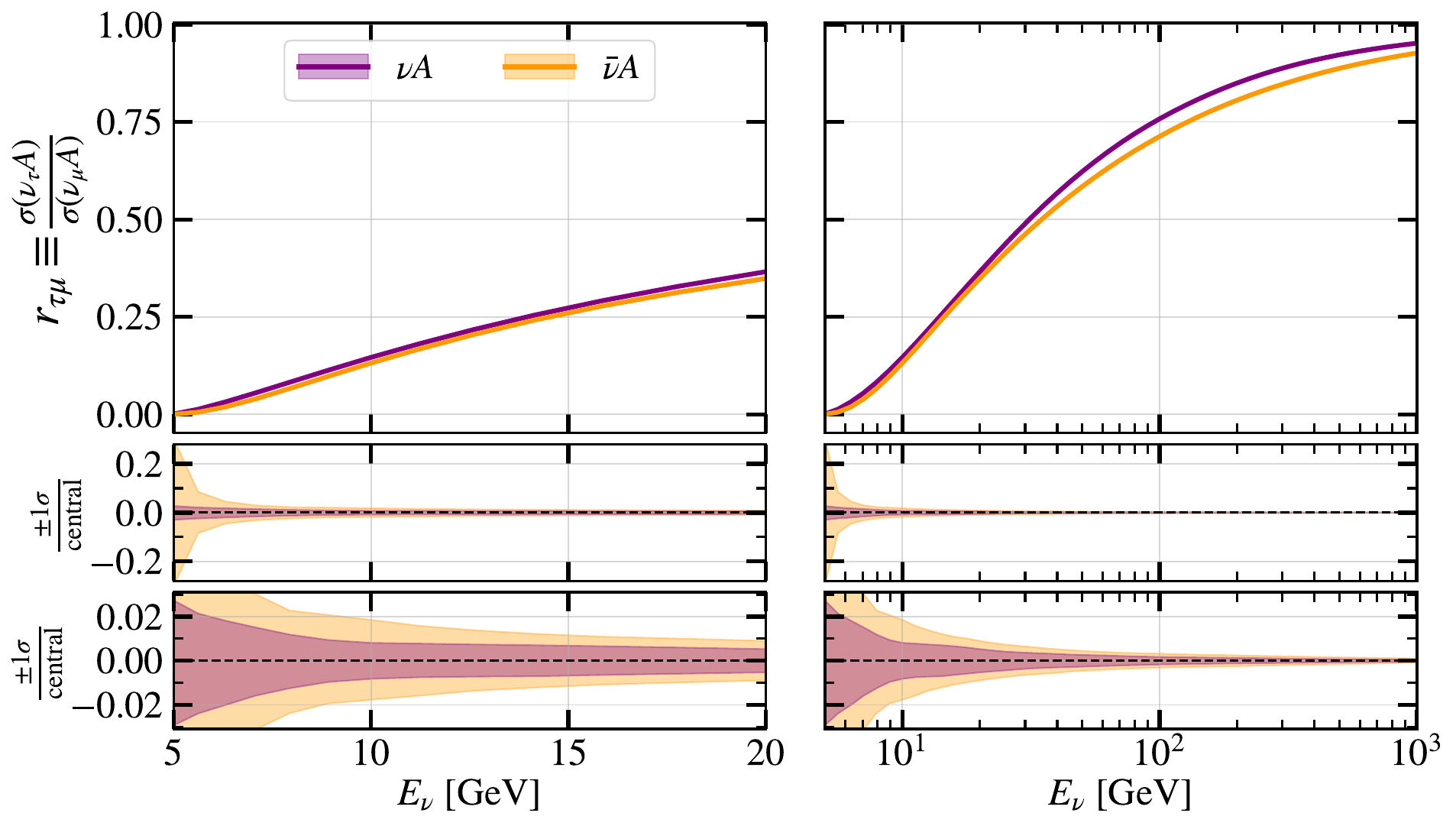}
\caption{Ratio of the $\nu_{\tau}$ CC DIS cross section with an isoscalar target $A$ to that of $\nu_{\mu}$, shown as a function of neutrino energy $E_{\nu}$. The purple curves correspond to neutrinos, and the orange ones -- to antineutrinos. The middle and the bottom panels show the relative 68\% ($1\sigma$) uncertainty on the cross section ratios as derived from the NNPDF4.0 Monte Carlo replicas.}
\label{fig:ratio_uncertainty}
\end{figure*}

Having the individual replicas of the total $\nu_{\tau}A$ and $\nu_{\mu}A$ cross sections at each probed $E_{\nu}$, we build the distribution of the cross section ratios $r_{\tau\mu} \equiv \sigma(\nu_{\tau}A) / \sigma(\nu_{\mu}A)$ as a function of $E_{\nu}$ and construct the corresponding 68\% confidence levels contours. The results are given in \cref{fig:ratio_uncertainty}.

We find that the PDF-driven uncertainty on the $\sigma(\nu_{\tau}A) / \sigma(\nu_{\mu}A)$ cross section ratio is $\lesssim 3\%$ ($\lesssim 30\%$) at \SI{5}{GeV} and $\lesssim 1\%$ ($\lesssim 2\%$) at \SI{10}{GeV} for neutrinos (antineutrinos). The uncertainty monotonically decreases as a function of neutrino energy and reaches sub-percent level for both neutrinos and antineutrinos above \SI{20}{GeV}. Whether the PDF-driven uncertainty on the $\nu_{\tau} / \nu_{\mu}$ cross section ratio is significant enough to be included in an experimental analysis is therefore dependent on the energy reach of the experiment and the sensitivity to the $\bar{\nu}_{\tau}$ events. In the specific case of atmospheric neutrinos, the $\bar{\nu}_{\tau}$ flux (resulting mainly from $\bar{\nu}_{\mu} \to \bar{\nu}_{\tau}$ oscillations in the energy range from a few GeV to a few hundreds of GeV) is subdominant compared to the ${\nu}_{\tau}$ flux. Additionally, the $\bar{\nu}_{\tau}$ CC DIS cross section is $\sim$\SIrange[range-phrase=--,range-units=single]{2}{10}{} times smaller at these energies than that of $\bar{\nu}_{\tau}$. As the result, one could expect $\sim$4 times fewer tau antineutrinos than neutrinos detected in an experiment such as IceCube in the \SIrange[range-phrase=--,range-units=single]{5}{1000}{GeV} energy range\footnote{This estimate is derived from the ratio of the expected $\nu_{\tau}$ and $\bar{\nu}_{\tau}$ $\{$flux $\times$ cross section$\}$ products integrated over the specified energy range. The atmospheric neutrino fluxes for $\nu_{\mu}$ and $\bar{\nu}_{\mu}$ are calculated via \textsc{MCEq} \cite{Fedynitch:2015zma,Fedynitch:2018cbl} with the \textsc{DPMJet-III} 19.1 hadronic interaction model and oscillated into $\nu_{\tau}$ and $\bar{\nu}_{\tau}$ using the \textsc{Neurthino} oscillation probability code \cite{Neurthino:2021xxx}. The energy-dependent detector acceptance is not taken into account.}. The expected $\nu_{\tau} / \bar{\nu}_{\tau}$ event ratio further increases with decreasing energy.  Therefore, given that only the antineutrinos with $E_{\nu} \lesssim\,\,$\SI{7}{GeV}  have a non-negligible tau-to-muon CC DIS cross section ratio uncertainty ($>$5\%), and that their contribution to the overall $(\nu_{\tau} + \bar{\nu}_{\tau})$-CC event rate is small, our recommendation is for the corresponding systematic uncertainty to \textit{not} be included in the atmospheric neutrino oscillation analyses. However, the explicit consideration of the PDF-driven uncertainty on the $\bar{\nu}_{\tau} / \bar{\nu}_{\mu}$ CC DIS cross section ratio is generally advisable for beam experiments capable of operating in an antineutrino mode and detecting large samples of $\bar{\nu}_{\tau}$ CC events at energies below \SI{10}{GeV}. While this could in principle apply to DUNE, only $\sim$40 $\bar{\nu}_{\tau}$ CC events are projected for 3.5 years of exposure in the antineutrino mode, including DIS, RES, and QE interactions \cite{DeGouvea:2019kea}. This implies that the initial $\bar{\nu}_{\tau}$ DIS samples from DUNE will be statistics-limited and likely insensitive to the PDF-driven $\sim$\SIrange[range-phrase=--,range-units=single]{2}{30}{}$\%$ uncertainties on the $\bar{\nu}_{\tau} / \bar{\nu}_{\mu}$ DIS cross section ratio. 

\section{Impact of the invariant hadronic mass threshold}\label{sec:pdf_wmin_impact}

When evaluating the differential cross section $\frac{\mathrm{d}\sigma}{\mathrm{d}y}$ in \cref{eq:dsigma_dy}, we imposed a cut on the invariant mass of the hadronic final state, requiring it to be larger than $W_{\mathrm{min}} =\,$\SI{1.4}{GeV} for the interaction to be classified as deep inelastic scattering. However, this setting of $W_{\mathrm{min}}$ is not universal and depends on the analyzer's choice of the resonances to be included in the deep inelastic scattering region \cite{Jeong:2023hwe}. In the \textsc{genie} event generator, the default setting is $W_{\mathrm{min}}=\,$\SI{1.7}{GeV}\footnote{Note that in \textsc{genie}, the equivalent variable is called $W_{\mathrm{cut}}$, while $W_{\mathrm{min}} \equiv M_{N} + m_{\pi}$, with $m_{\pi}$ being the pion mass, corresponds to the lower end of the RES energy range.}\cite{Andreopoulos:2009rq}. It is therefore natural to test what impact the choice of $W_{\mathrm{min}}$ makes on the $r_{\tau\mu}$ cross section ratio and compare it to the PDF-driven uncertainty. 

\begin{figure*}[htb!]
\includegraphics[width=0.7\textwidth]{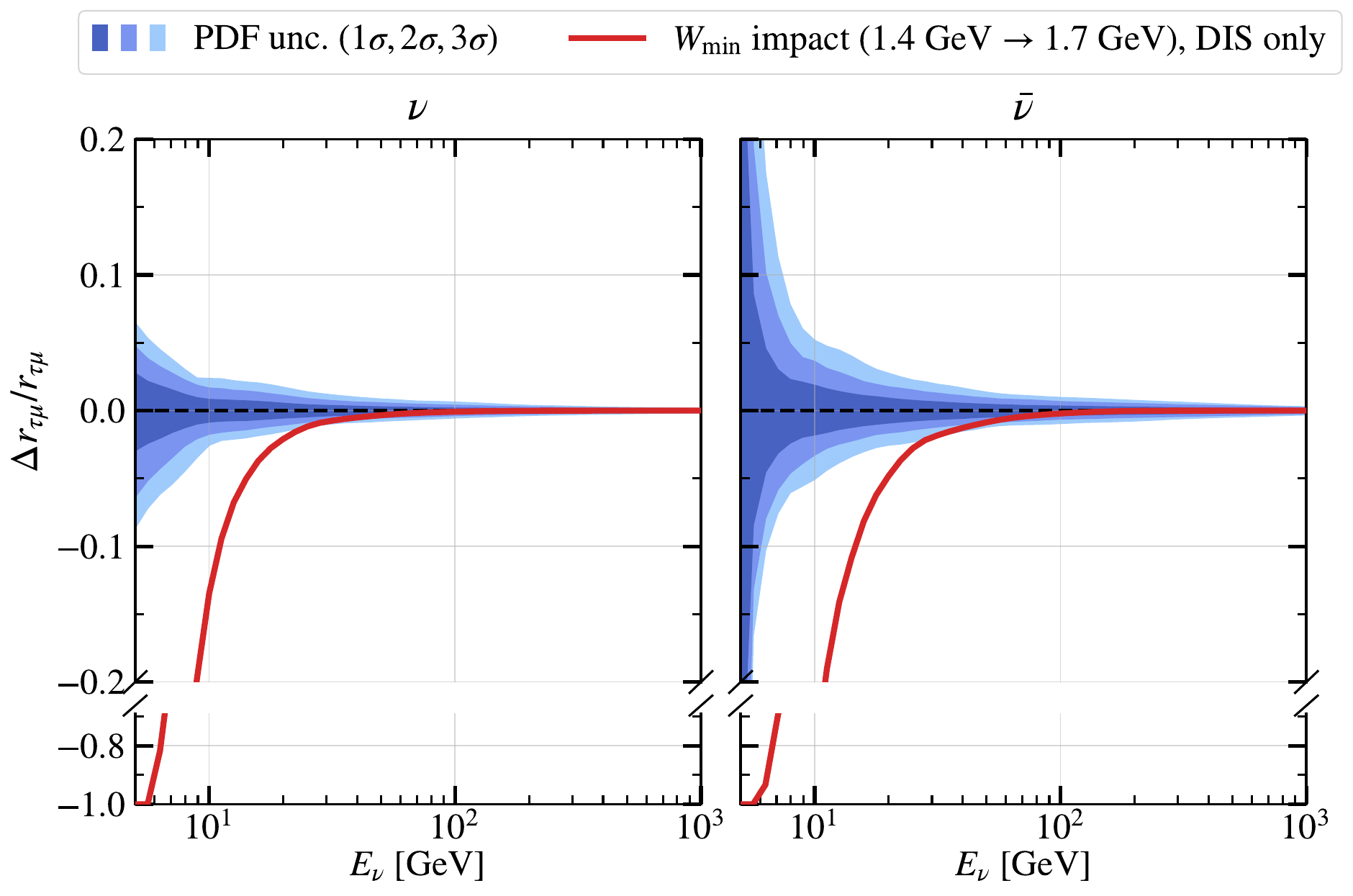}
\caption{Red curve: relative impact of a naive shift in the invariant hadronic mass threshold ($W_{\mathrm{min}}$) from \SI{1.4}{GeV} to \SI{1.7}{GeV} on the $\nu_{\tau}$/$\nu_{\mu}$ CC DIS cross section ratio ($r_{\tau\mu}$), disregarding the corresponding impact on the resonance cross section. Blue bands: relative impact of the NNPDF uncertainties on $r_{\tau\mu}$, as derived from the NNPDF Monte Carlo replicas. The $y$-axis is broken to show the full impact of $W_{\mathrm{min}}$ at the lowest considered neutrino energies. This figure serves only as an illustration of the $W_{\mathrm{min}}$ impact if not compensated by other parameters of the DIS and RES cross sections (see text for details).}
\label{fig:wmin_impact_analytical}
\end{figure*}

\begin{figure}[htb!]
\includegraphics[width=0.85\textwidth]{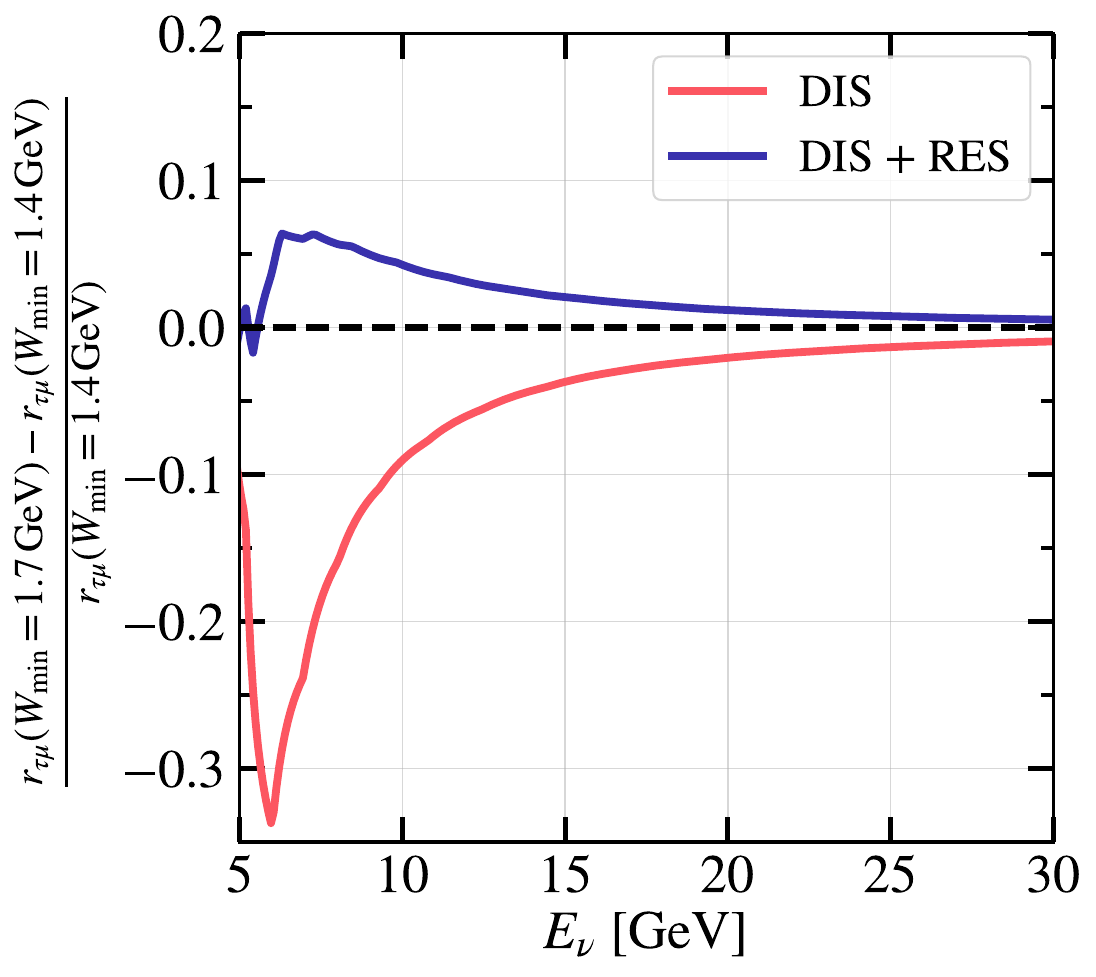}
\caption{Relative difference between the $\nu_{\tau}$/$\nu_{\mu}$ charged current cross section ratios $r_{\tau\mu}$ evaluated with the invariant hadronic mass thresholds $W_{\mathrm{min}} = 1.4\,\mathrm{GeV}$ and $W_{\mathrm{min}} = 1.7\,\mathrm{GeV}$. The red curve shows the difference in the ratios of the CC DIS cross sections, and the dark blue curve -- CC DIS + RES cross sections. All cross sections are evaluated in \textsc{genie} 2.12.8 \cite{Andreopoulos:2009rq} with the Bodek-Yang-corrected GRV98 PDF at LO \cite{Bodek:2002ps,Gluck:1998xa}.}
\label{fig:wmin_impact_genie}
\end{figure}

In \cref{fig:wmin_impact_analytical}, we show the relative impact of the PDF variation within its \SIrange[range-phrase=--,range-units=single]{1}{3}{}$\sigma$ uncertainty limits as well as the impact of changing $W_{\mathrm{min}}$ from \SI{1.4}{GeV} to \SI{1.7}{GeV}. We find that at $E_{\nu} \lesssim 30\,\mathrm{GeV}$, the difference between the tau-to-muon CC DIS cross section ratios obtained with the two different choices of $W_{\mathrm{min}}$ is not contained in the $3\sigma$ PDF uncertainty bounds, and grows to nearly 100\% at \SI{5}{GeV} for both $\nu$ and $\bar{\nu}$. While these differences appear to be significant at first, one must remember that drop in the DIS cross section due to a higher $W_{\mathrm{min}}$ threshold is accompanied by the increase in the RES cross section. In the \textsc{genie} framework, this is handled by the RES-DIS joining scheme, the details of which can be found in \cite{Andreopoulos:2009rq}. In \cref{fig:wmin_impact_genie}, we zoom into the \SIrange[range-phrase=--,range-units=single]{5}{30}{GeV} energy region and show the impact on the DIS, and DIS + RES $\nu_{\tau}$/$\nu_{\mu}$ cross section ratios due to the shift in $W_{\mathrm{min}}$, as evaluated within \textsc{genie}. As expected, the impact on the ratios of the DIS+RES $\nu_{\tau}$/$\nu_{\mu}$ cross sections is much smaller (reaching at most 6\%) than the impact on the DIS cross section ratio alone (reaching \SIrange[range-phrase=--,range-units=single]{10}{30}{}\% at $E_{\nu} \lesssim 10\,\mathrm{GeV}$). This implies that if a systematic uncertainty on the $r_{\tau\mu}$ ratio due to the arbitrariness in the choice of $W_{\mathrm{min}}$ is to be included in an analysis, it must be implemented for both DIS and RES channels to ensure a physical transition between the two. 

Furthermore, the cross section in the RES-DIS transition region is constrained by inclusive and exclusive cross section data from the bubble chamber experiments such as ANL 12FT \cite{Radecky:1981fn,Day:1983itd}, BNL
7FT \cite{Fanourakis:1980si,Kitagaki:1986ct,Baker:1981su,Baker:1982ty}, BEBC \cite{BEBCTSTNeutrino:1983vyc,Colley:1979rt,Aachen-Bonn-CERN-London-Oxford-Saclay:1977yve,Aachen-Bonn-CERN-Democritos-London-Oxford-Saclay:1981nog,Aachen-Bonn-CERN-Munich-Oxford:1980iaz,Aachen-Birmingham-Bonn-CERN-London-Munich-Oxford:1985yec,Allasia:1990uy,Bell:1978qu}, and FNAL 15FT \cite{Baker:1982jf,Asratian:1984ir,Taylor:1983qj}. The $W_{\mathrm{min}}$ setting therefore cannot be varied on its own without the corresponding changes in the other parameters of the RES and the DIS models. In particular, $W_{\mathrm{min}}$ is anti-correlated with the $R^{\mathrm{CC}2\pi}$ coefficients modulating the contribution of the two-pion final states, as well as the axial mass $M_A^{\mathrm{RES}}$ entering the form factors of the resonant scattering cross sections \cite{NuSTEC:2019lqd,GENIE:2021zuu}. The most accurate approach would therefore be to keep $W_{\mathrm{min}}$ fixed, tune the rest of the RES and the DIS parameters to the cross section data, and determine their covariance matrix as done e.g. in the tuning efforts of the GENIE Collaboration \cite{GENIE:2021zuu}. One would then keep the tune parameters free in the experimental analysis in question and introduce a penalty term through the covariance matrix. With this approach, the uncertainty on the $r_{\tau\mu}$ ratio would be automatically handled across all interaction channels, while the introduction of the $r_{\tau\mu}$ uncertainty due to $W_{\mathrm{min}}$ alone would not be complete and might lead to unphysical fit outcomes.

\section{Summary and outlook}

In this study, we have considered the charged current deep inelastic scattering cross sections of muon and tau neutrinos and evaluated the uncertainty on their ratio at leading order in perturbation theory. This result was obtained by propagating the uncertainty on the parton distribution functions, for which we utilized the NNPDF Monte Carlo replicas, through the standard DIS calculation at LO. We found this uncertainty to be smaller than $3\%$ for neutrinos with energies above \SI{5}{GeV} and antineutrinos with  energies above \SI{10}{GeV}. Between \SI{5}{GeV} and \SI{10}{GeV}, the PDF-driven uncertainty on the antineutrino cross section ratio was found to vary between \SIrange[range-phrase=--,range-units=single]{2}{30}{}$\%$.  For the purpose of atmospheric $\nu_{\mu} \to \nu_{\tau}$ oscillation analyses with experiments such as IceCube-DeepCore and KM3NeT, these uncertainties can be considered negligible, as the atmospheric tau neutrino samples are dominated by $\nu_{\tau}$ rather than $\bar{\nu}_{\tau}$. We therefore conclude that in atmospheric neutrino analyses, the $\nu_{\tau}$ and the $\nu_{\mu}$ DIS cross section uncertainties should be coupled through a single parameter instead of being represented by two independent parameters. For the beam experiments with an antineutrino mode, accounting for the PDF-driven uncertainty on the $\bar{\nu}_{\tau} / \bar{\nu}_{\mu}$ cross section ratio may be worthwhile. However, in the case of DUNE, we expect that the DIS event rate uncertainty in the $\gtrsim$ \SI{5}{GeV} region will be predominantly statistical in the first few years of exposure \cite{DeGouvea:2019kea}.  

The smallness of the derived cross section ratio uncertainty stems from the correlations among the $F_i$ structure functions at the PDF level. In particular, although the contribution of the $F_5$ structure function to the DIS cross section is modulated by the lepton mass, $F_5$ itself is strongly correlated with $F_2$ and $F_1$. This is reflected in the Albright-Jarlskog and the Callan-Gross relations at LO. This implies that one cannot allow for additional freedom in the tau neutrino CC DIS cross section through $F_5$, as varying $F_5$ must lead to the corresponding variations in $F_1$ and $F_2$. Any physical variations in $F_5$ therefore modify both tau and muon neutrino cross sections and leave their ratio largely unaffected.

We expect these conclusions to be stable with respect to the order in perturbation theory at which the calculation of the structure functions is performed, given that the correlations between $F_1$, $F_2$, and $F_5$ are well preserved at NLO \cite{Kretzer:2002fr}. However, this could be more rigorously confirmed by utilizing modern cross section calculation tools such as \textsc{yadism} \cite{Candido:2024rkr}, which can be used up to N$^{3}$LO. Although the $F_4$ and $F_5$ structure functions -- and therefore, the tau neutrino cross sections -- are not implemented in the \textsc{yadism} framework at the time of writing, making such functionality available would be a natural next step for this study.

Finally, we comment on the impact of the invariant hadronic mass threshold, $W_{\mathrm{min}}$, on the DIS cross section ratio. We found that a naive shift in the threshold by \SI{0.3}{GeV} introduces differences of $\mathcal{O}$(\SIrange[range-phrase=--,range-units=single]{10}{100}{}\%) in the $\nu_{\tau}$/$\nu_{\mu}$ cross section ratio at neutrino energies below \SI{10}{GeV}, which is not covered by the $3\sigma$ range of the PDF-driven uncertainty. However, any shift in the DIS cross section due to a different choice of $W_{\mathrm{min}}$ must be accordingly reflected in the resonant scattering channels with $W < W_{\mathrm{min}}$. We find that the ratio of the DIS+resonance cross sections is affected at a much smaller level ($\sim$5$\%$) by the considered shift in $W_{\mathrm{min}}$ than the DIS cross section alone. Therefore, if the flexibility in the choice of $W_{\mathrm{min}}$ is to be incorporated as a source of the $\nu_{\tau}$/$\nu_{\mu}$ cross section ratio uncertainty, it must be done for both DIS and resonance cross sections in a correlated fashion. Furthermore, the correlations between $W_{\mathrm{min}}$ and other free parameters of the resonant and the deep inelastic scattering must be taken into account to ensure that the total cross section in the shallow inelastic scattering region agrees well with the bubble chamber data. This could be achieved by implementing a covariance matrix between the cross section parameters from one of the publicly available \textsc{genie} tunes \cite{GENIE:2021zuu} or, alternatively, by producing a new covariance matrix corresponding to a different fixed $W_{\mathrm{min}}$ used in the analysis in question. These efforts might be relevant for the experiments having a significant portion of the shallow inelastic scattering events, such as IceCube-Upgrade and DUNE.

\section*{Acknowledgements}
The authors acknowledge the support from the Carlsberg Foundation (project no. 117238) and thank Mary Hall Reno for valuable feedback on the manuscript.

\clearpage
\appendix

\counterwithin{figure}{section}
\setcounter{table}{0}
\renewcommand{\thetable}{A\arabic{table}}

\onecolumngrid

\section{Impact of the charm quark and the target mass corrections}

\begin{figure*}[htb!]
\includegraphics[width=0.7\textwidth]{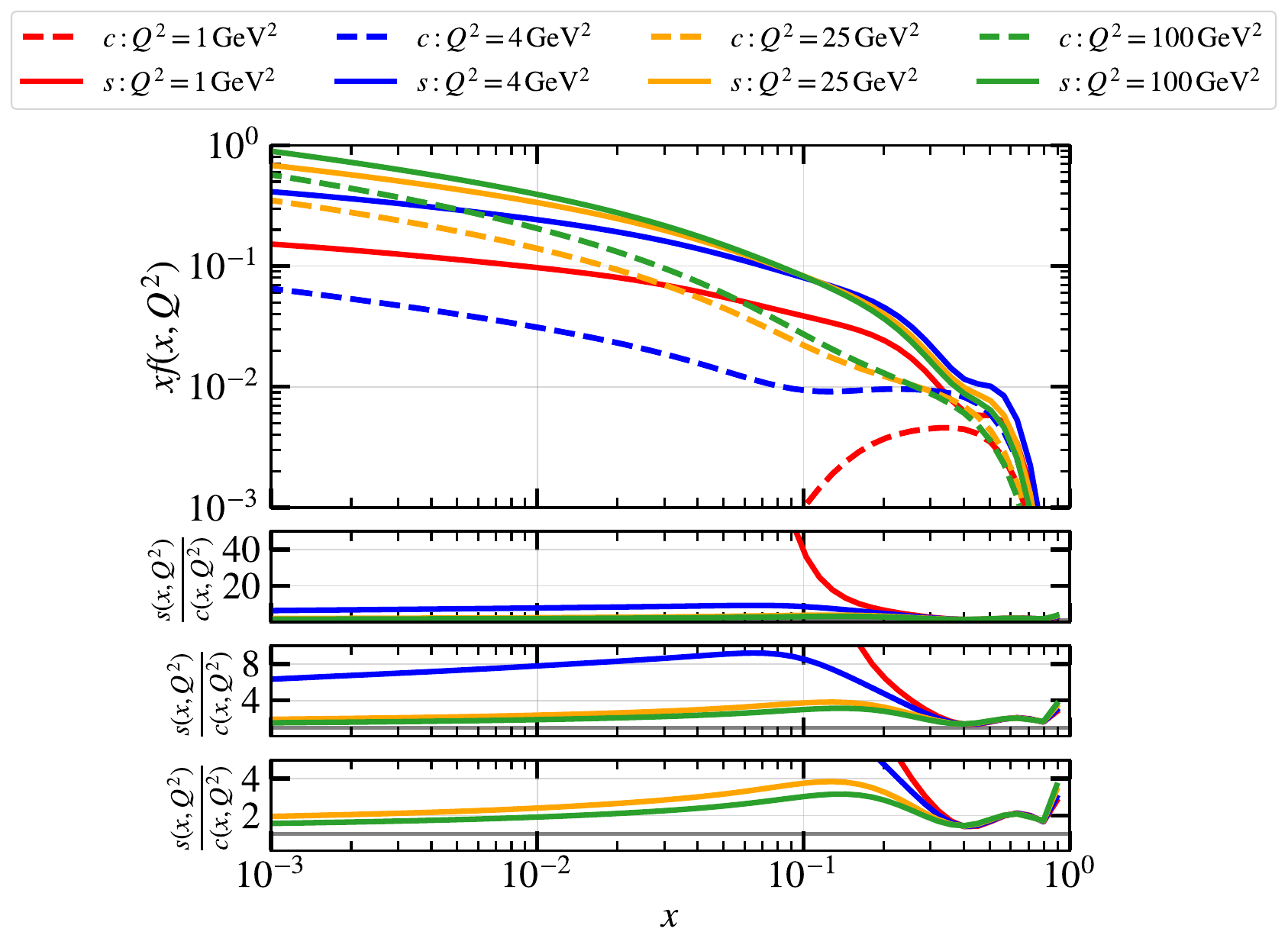}
\caption{\textit{First row}: densities of the sea strange and charm quarks inside inside a proton, shown at several fixed values of $Q^2$ as a function of Bjorken $x$. \textit{Second to fourth rows}: ratio of the strange quark density to that of the charm quark (shown for three different sets of the $y$-axis limits for clarity). The displayed PDFs correspond to average of the 1000 NNPDF4.0 Monte Carlo replicas (0th members of the PDF set) \cite{NNPDF:2021njg}.}
\label{fig:sea_charm_density}
\end{figure*}

\begin{figure*}[htb!]
\includegraphics[width=0.7\textwidth]{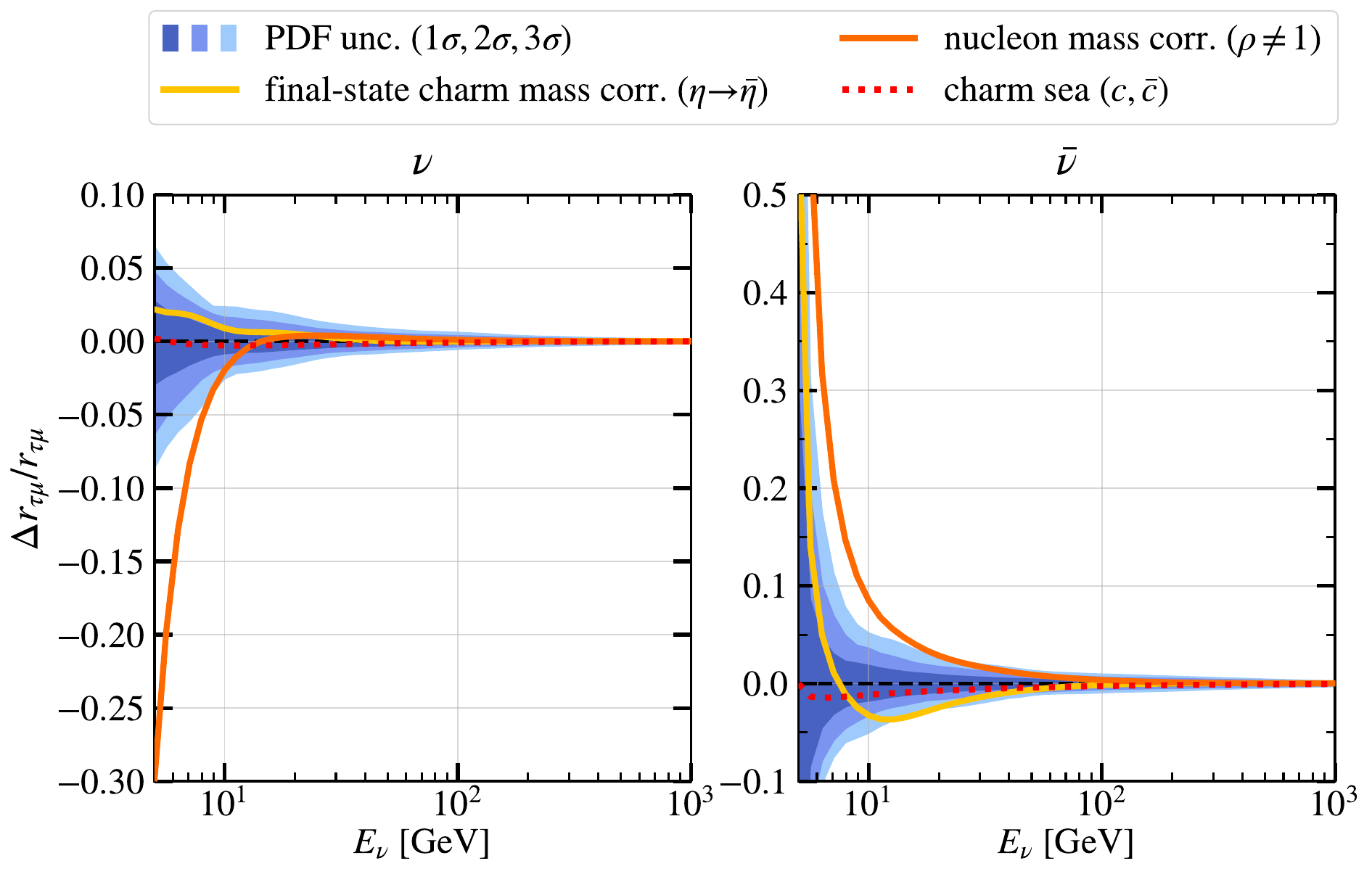}
\caption{Relative impact on the $\nu_{\tau}/\nu_{\mu}$ CC DIS cross section ratio of the final-state charm mass correction as per \cref{eq:nachtmann_eta_correction_quark_mass} (solid yellow curve); the target nucleon mass corrections as per \cref{eq:rho_definition,eq:proton_Fi_at_LO_corrected_by_MN_mc} (solid orange curve); the inclusion of the sea charm quark densities into the structure function calculations (dotted red curve); and the PDF uncertainties derived from the NNPDF Monte Carlo replicas \cite{NNPDF:2021njg} (blue bands).}
\label{fig:charm_tmc_impact_on_xsec_ratio}
\end{figure*}

\twocolumngrid

\bibliographystyle{apsrev4-1}
\bibliography{main.bib}

\end{document}